\documentclass[12pt, letterpaper]{article}
\usepackage{graphicx}
\usepackage{color}
\usepackage{wrapfig}
\usepackage{url}
\usepackage{hyperref}
\usepackage{amssymb}

\newcommand{\YuH}[1]{{\color{black}{#1}}}

\begin{document}

\title{NETWORK OF SCIENTIFIC CONCEPTS: EMPIRICAL ANALYSIS AND MODELING\footnote{Preprint of an article accepted for publication in  Advances in Complex Systems journal [copyright World Scientific Publishing Company] 
}}

\author{VASYL PALCHYKOV$^1$ \footnote{palchykov@gmail.com} \and MARIANA KRASNYTSKA$^{1,2}$  \and OLESYA MRYGLOD$^{1,2}$ \and YURIJ HOLOVATCH$^{1,2,3}$}

\maketitle

$^1$ L$^4$ Collaboration \& Doctoral College for the Statistical Physics of Complex Systems,
Leipzig-Lorraine-Lviv-Coventry, Europe \\
$^2$  Institute for Condensed Matter  Physics, National Acad. Sci. of Ukraine, UA--79011 Lviv, Ukraine \\
$^3$ Centre for Fluid and Complex Systems, Coventry University, Coventry,
CV1 5FB, United Kingdom


\begin{abstract}

Concepts in a certain domain of science are linked via
intrinsic connections reflecting the structure of knowledge.
To get a qualitative insight and a quantitative description
of this structure, we perform empirical analysis and modeling
of the network of scientific concepts in the domain
of physics. To this end we use a collection of manuscripts submitted 
to the e-print repository \texttt{arXiv}  and the vocabulary of 
scientific concepts collected  via the \texttt{ScienceWISE.info} 
platform and construct a network of scientific concepts based on 
their co-occurrences in publications.  The resulting complex network
possesses a 
number of specific features (high node density, dissortativity, structural
correlations, skewed node degree distribution) that can not be understood as a result
of simple growth by several commonly used network models. We show that the 
model based on a simultaneous account of two factors, growth by blocks and preferential 
selection, gives an explanation of empirically observed properties of the concepts network. 
\end{abstract}

Keywords: logology, complex systems, semantic networks, concepts network,
generative models

\section{Introduction}\label{I}

Semantic networks, i.e., networks that reflect connections between concepts within a particular domain, are among the tools to formalize knowledge as a whole \cite{SowaEncyclopedie}. Their history (at least their written history) can be traced back to the famous Porphyrian
tree\footnote{A diagram representing the classification of substances in Aristotle’s ``Categories'' (in a written, not drawn form though) by Neoplatonist philosopher Porphyry\cite{Horrocks2008}.} and extended up to the modern ontologies in a computer and information science \cite{Fontoura2006,Fensel2001}.
The description of knowledge system is also a part of logology –- the science of science,
that aims in quantitative understanding of origins of scientific discovery and creativity, its structure and practice \cite{Zeng17,Wang21}. Since scientific publication remains the main form of documentation of the research output, the structure of scientific domains can be mapped using the vast amounts of currently available bibliographic data \cite{Handbook2019}.

While it is impossible to know the exact structure of the abstract system of knowledge, there are many ways to model it using its projections. The semantic space for different domains can be modeled as a complex network of topic-indicating labels. 
To name a few, one can mention analysis of the topical landscape for research papers  based on the network of co-used PACS (Physics and Astronomy Classification Scheme) numbers \cite{Herrera2010,Pan2012}; co-mentions of chemical entities in biomedical papers \cite{Foster2015}; close co-occurrence of pre-defined terms in full texts of papers in cognitive neuroscience \cite{Beam2014}; co-appearance of pre-defined concepts in titles or abstracts of papers in quantum physics \cite{Krenn2020}; analysis of mutual hyperlinks between Wikipedia pages devoted to mathematical theorems \cite{Silva2010}. The similarities between documents, authors, research groups, and other entities can be established using information about co-occurrence of terms, co-usage of keywords or topical indices, co-citations or bibliographic couplings, etc. Complex network formalism allows us to visualize \cite{vanEck2009}  and quantify the structure of such similarities, which are typically considered as indicators of topical relatedness and, therefore, as projections of knowledge.

The networks discussed emerge as an outcome of a dynamic process at which the new knowledge is acquired. The new pieces of knowledge or innovations can be modeled either as an emergence of new ideas or concepts (i.e. new nodes in network representation) or new connections between the existing ones (i.e. by establishing new or reinforcing existing network links) \cite{Iacopini2018,Uzzi2013,Palchykov20}. Modeling such processes is a challenging task both for its fundamental relevance and numerous practical implementations.
It may be used to build an efficient policy aimed at financial support or targeted stimulation of national research or to detect ``hot topics'' and emerging trends for research topic selection. The quantitative analysis of scholarly metadata provides a possibility to reveal not only explicit interrelations and patterns but also implicit ones \cite{Evans2011}.


The process of scientific discovery is governed by the structure of scientific knowledge, but at the same time the evolution of this structure is dependent on the process of appearing new concepts and links between them. Such interplay between the structure and dynamics
is a typical feature of any complex system \cite{Thurner18,Holovatch17}.
Sometimes it is hard to distinguish the evolution of the principal structure of knowledge from the dynamic process governed by this structure since both are recursively interrelated. To this end, the models are constructed to capture how terms, keywords, labels, or tags become co-chosen from some predefined semantic space \cite{Cattuto2009,Iacopini2018,Rzhetsky2015}.

\begin{figure}[h]
 \centerline{\includegraphics[width=13cm]{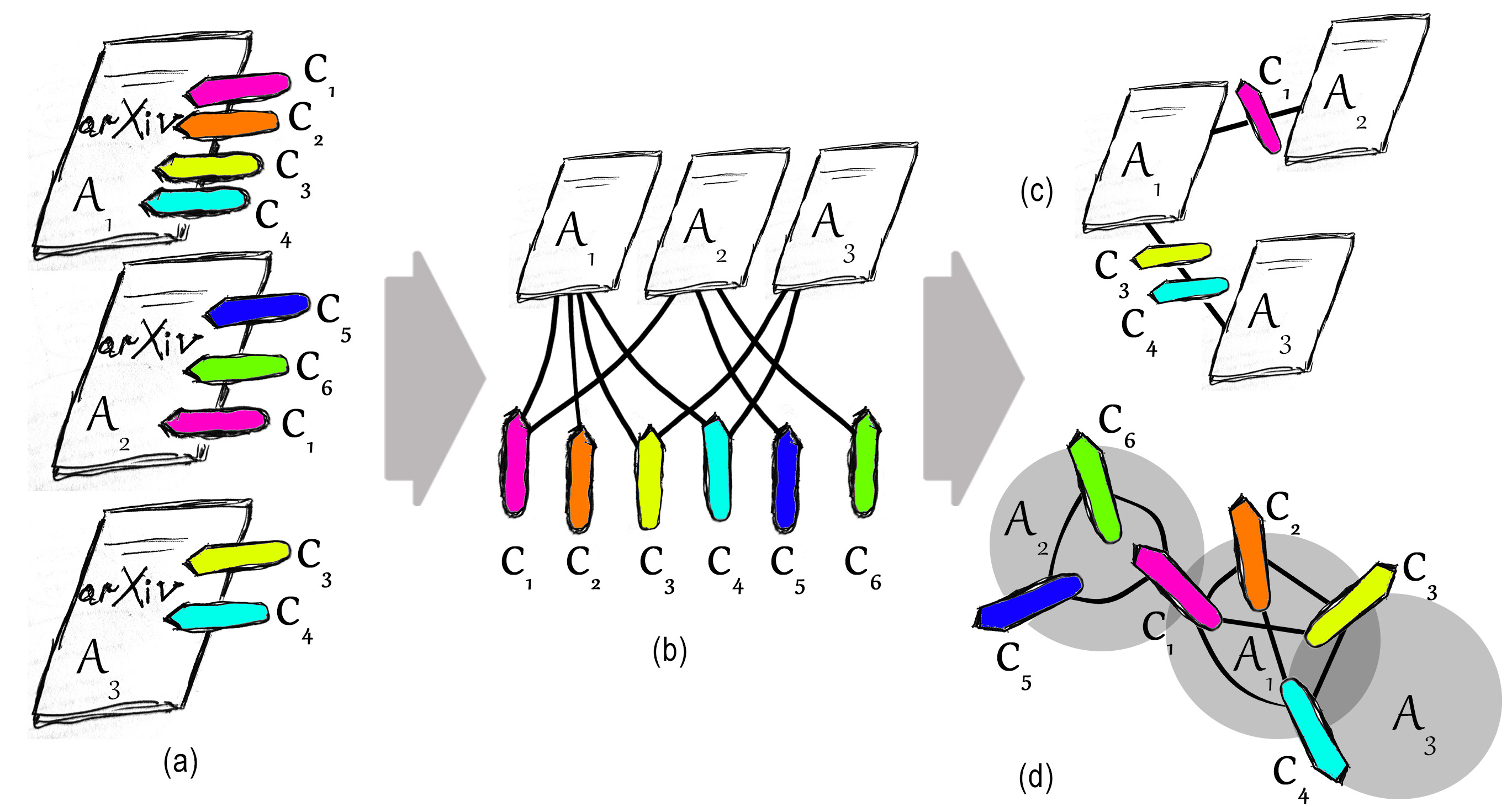}}
 \caption{Construction of the network of scientific concepts analysed in this paper.
 Articles $A_i$ containing certain number of different concepts $C_j$ (panel {\bf a}) are represented in a form of a bipartite network that consists of nodes of two types: articles $A_1,A_2,\dots,A_{\cal N}$
 and concepts $C_1,C_2,\dots,C_N$ (panel {\bf b}). In the bipartite network, links are present only between nodes of different types: each $A_i$ node is linked to the $C_j$ node representing a concept encountered in this article. Single-mode projections of the bipartite network result in the network of articles (panel {\bf c}) and of concepts (panel {\bf d}). The data considered in our paper is based on ${\cal N}=36,386$ \texttt{arXiv.org} articles and contains $N=11,853$ unique concepts.
  \label{fig:data}}
\end{figure}

With all said above, it is of a primary importance to analyze the structure of semantic networks in science and to model processes of their evolution. In this paper, we plan to address both questions. To this end, we will use a collection of manuscripts submitted to the e-print repository \texttt{arXiv.org} and the vocabulary of scientific concepts collected via the 
\texttt{ScienceWISE.info} platform\footnote{\texttt{ScienceWISE.info} is a web service connected to the main online repositories such as \texttt{arXiv}, whose peculiarity is a bottom-up approach in
the management of scientific concepts (\url{http://sciencewise.info/faq}). 
The concepts are extracted automatically from scientific publications using \texttt{KPEX} algorithm \cite{kpex} and then validated by registered users of the platform, see also \cite{palchykov2016} for more details.}, 
see Fig. \ref{fig:data}. 
To our knowledge, this vocabulary currently contains the most comprehensive collection of scientific concepts in the domain of physics. 
To consider relations between the concepts, we will analyze how do the concepts co-occur in publications. Such analysis will allow constructing the network of concepts -- empirical analysis and modeling of this network is our main subject of study in this paper. Some of our results were previously announced in a Letter \cite{Palchykov21}.

The rest of the paper is organized as follows. In the next section we describe the data under analysis in detail, explain how the network of 
scientific concepts is constructed, and analyze its essential features. 
We show that the empirically observed network is a dense one and possesses other non-trivial characteristics that cannot be understood within simple 
generative models \YuH{as Erd\H{o}s-R\'enyi random graph (section \ref{II.3}) or Barab\'asi-Albert model (section \ref{II.4})}. To understand possible mechanisms that lead to the concepts network under discussion, in Section \ref{III} we develop a model 
that reproduces the principal empirically observed features. The key assumption of the model is growth by blocks with preferential selection. 
We show that both ingredients are crucial in network modeling and further discuss the relation of the suggested model to the others used in 
dense networks modeling. We discuss the results in Section~\ref{IV} and finish by conclusions and outlook in Section~\ref{V}.

\section{Scientific concepts network}\label{II}

\subsection{Data}\label{II.1}
We consider a collection of scientific publications in Physics domain. The sample of manuscripts consists of 36,386 preprints submitted to e-repository \texttt{arXiv.org} during a single year 2013 that have been assigned to a single category during the submission process. Cross-categorical articles have been discarded from the current analysis in order to have one-to-one correspondence with the data sets  analyzed in \cite{palchykov2016,Palchykov18,Palchykov20}. For each of the manuscripts, a set of its inherent concepts has been extracted using \texttt{ScienceWISE} \cite{ScienceWISE} platform. 
In this way, the data under consideration, see Fig.~\ref{fig:data} (a),  can be conveniently described in the form of a bipartite network (this and other definitions related to the theory of complex networks in more details can be found, e.g., in \cite{barabasi2014network}) that consists of nodes of two types: articles $A_1,A_2,\dots,A_{\cal N}$ and concepts $C_1,C_2,\dots,C_N$, see Fig. \ref{fig:data}~(b). In \cite{palchykov2016} the purpose was to analyse the structure of the single-mode projection of the bipartite network into the space of articles Fig.~\ref{fig:data} (c). In this way, the communities  has been found that reveal an inner thematic structure. In turn, the analysis
reported below will concern the network of concepts: being of interest {\em per se}, it may serve also as a 
complementary step in analysis of the whole bipartite network, Fig.~\ref{fig:data} (d), via its another single-mode projection.

\begin{figure}[th]
 \centerline{\includegraphics[width=9.0cm]{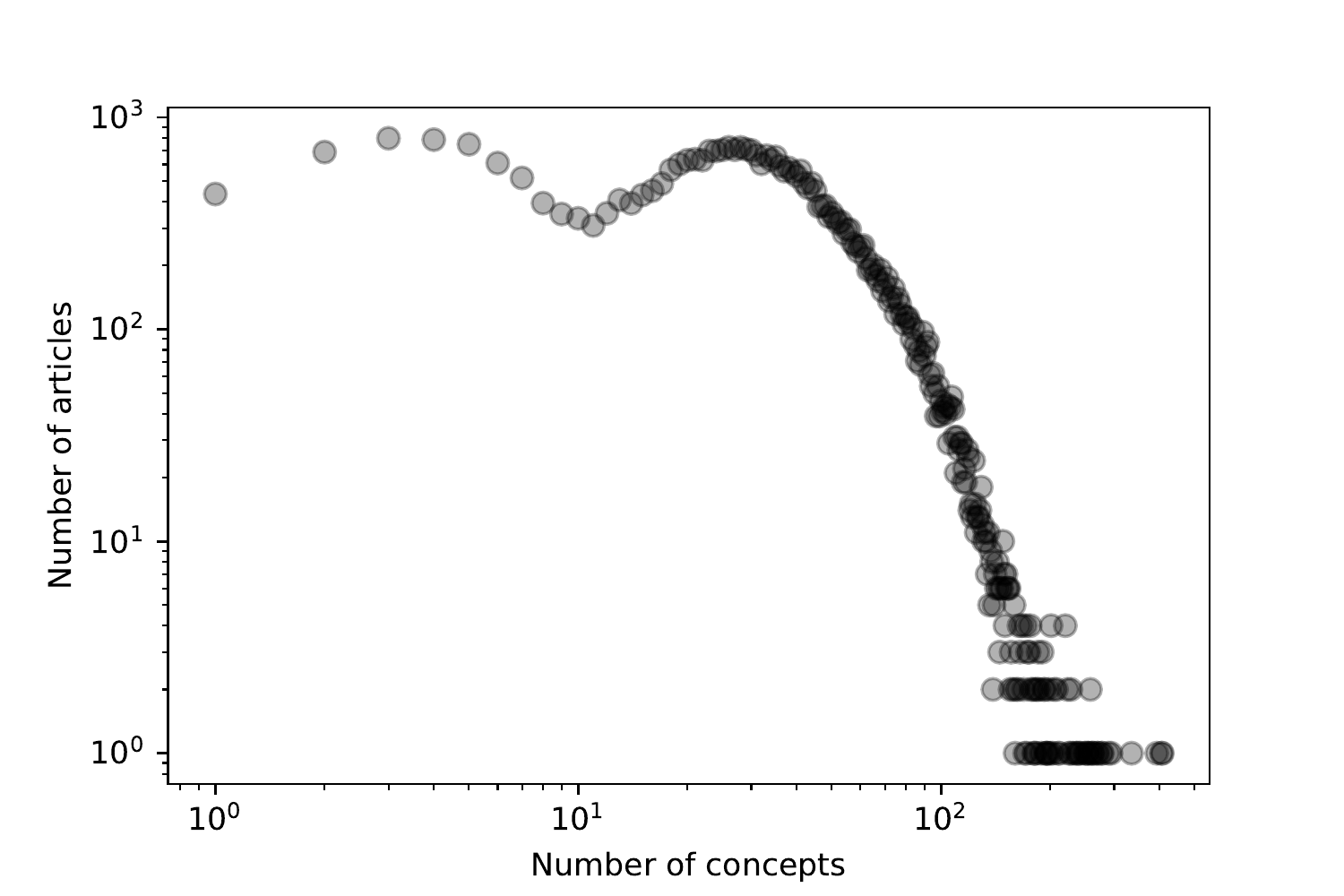}}
 \vspace*{8pt}
 \caption{Distribution of the number of concepts identified within the full text of an article. The distribution has a skewed shape with average of some $37$ concepts per article and the tail reaching over $400$ concepts per manuscript.}
 \label{fig:1.concepts_per_article}
\end{figure}

In the process of concept extraction, the \texttt{ScienceWISE} platform classifies concepts as generic and non-generic ones.\footnote{\YuH{See Appendix where we give several
examples of generic and non-generic scientific concepts as extracted via the ScienceWISE platform.}}   \YuH{Overall number of  concepts in the data set
under consideration 
is 12,200, out of these 347 concepts are generic and 11,853 are non-generic ones.}
In Ref.~\cite{palchykov2016}, generic concepts have been excluded from the analysis to avoid over-densifying the article network. 
To have a direct correspondence with the research conducted in \cite{palchykov2016}, the concepts marked as generic ones by \texttt{ScienceWISE} have 
been excluded from our analysis in this paper too. As a result of such routine we have found that a manuscript contains on average 37 concepts. The distribution of the number of concepts per manuscript has rather skewed shape as shown in Fig.~\ref{fig:1.concepts_per_article}, where the tail reaches about $400$ concepts per manuscript.
Beside a maximum around 30 concepts per article one may observe another one around 3 concepts, which may correspond to short report style submissions. Alternatively, the cause could lie in document parsing process: for some manuscripts \texttt{ScienceWISE} was not able to scan the full text, but could scan only abstract instead.

\subsection{Network construction and its basic features}\label{II.2}
The data set described contains $11,853$ distinct non-generic concepts.
We consider them as building blocks of the system of disciplinary  knowledge. However, not only ideas are 
important in scientific  creative processes, but the way how they are combined. Therefore, we represent each 
concept by a node of the network, and connect two nodes by a unit-weight link if the corresponding concepts 
appeared together within at least a single publication. The structure and \YuH{evolution} of this network,
called hereafter an \texttt{empirical} concepts network, is a subject of our analysis.

The resulting empirical concepts network consists of $N = 11,853$ nodes connected by $L = 5,382,448$ 
links\footnote{\YuH{All network properties were calculated using a \texttt{Python} implementation of 
\texttt{igraph} package \cite{igraph}.}}. All, except two isolated nodes, belong to the largest connected component. In a considered collection of scientific publications, each of concepts represented by the isolated nodes appeared only once in a single document. These concepts are \texttt{Finite strain theory} and \texttt{Nanomotor}. The manuscripts they appear in do not contain any other non-generic concepts (in fact, each of them contains 2 or 3 other generic concepts, which were excluded from consideration).

The density of links $\rho$, i.e. the ratio between the total number of links $L$ and the number of all possible links in the network $\rho = 2L/N(N-1)$, reaches $\rho=908/11852=7.66\%$.  
Such high value is not typical for networks actively analyzed in the literature. For example, density of such benchmark networks as Internet has $\rho=0.003\%$, $\rho=0.05\%$ for power grid, $\rho=0.02\%$ for scientific collaboration network, see e.g. \cite{barabasi2014network} and references therein.
As one of the consequences of high density of links, average node degree in this network is quite high as well: $\langle{k}\rangle=908$. In this respect, it is interesting to relate the network under consideration with the so-called dense networks, see e.g. \cite{Courtney18}. 

\begin{figure}[ht]
 \centerline{
\begin{tabular}{cc}
\includegraphics[width=6cm]{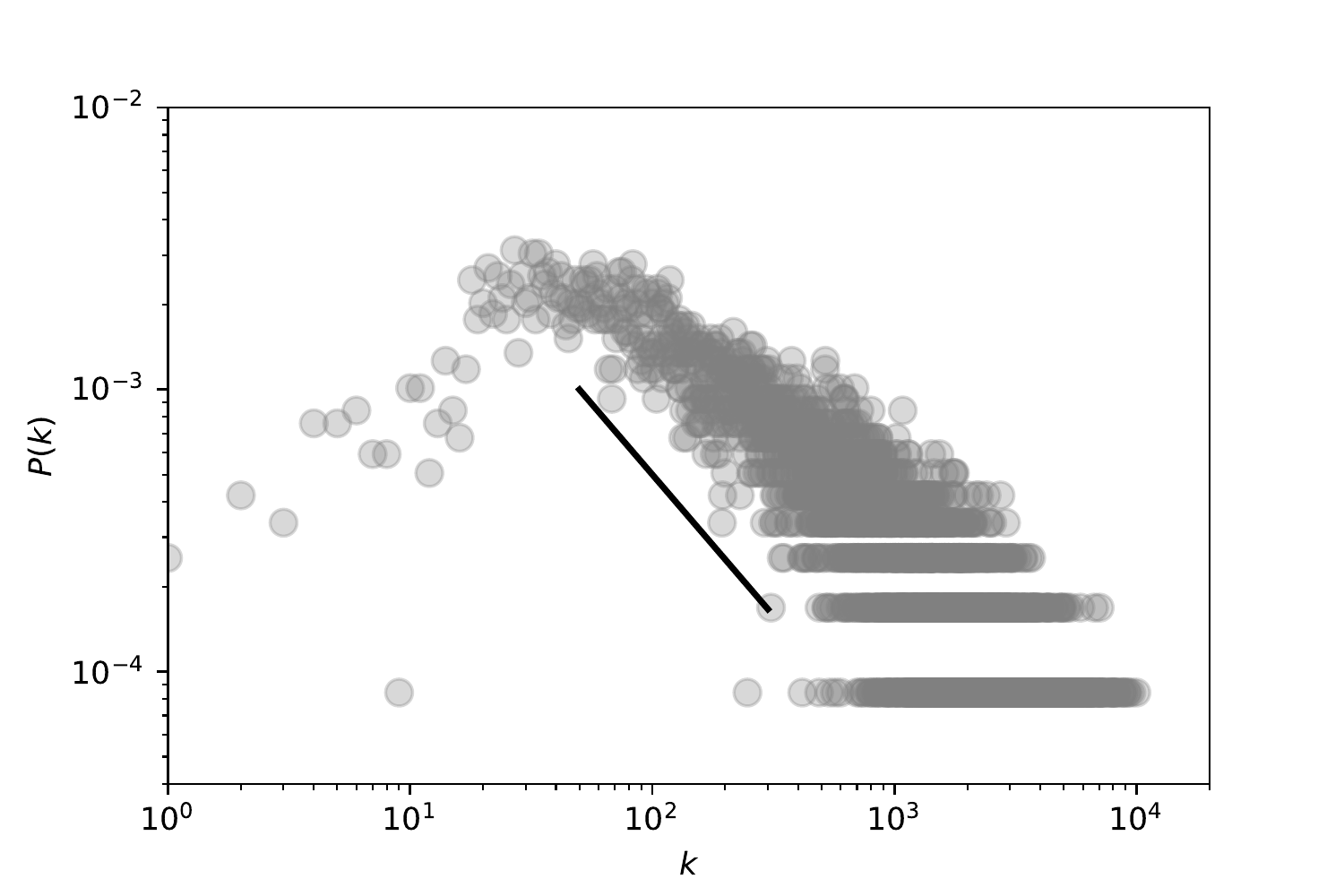} &
 \includegraphics[width=6cm]{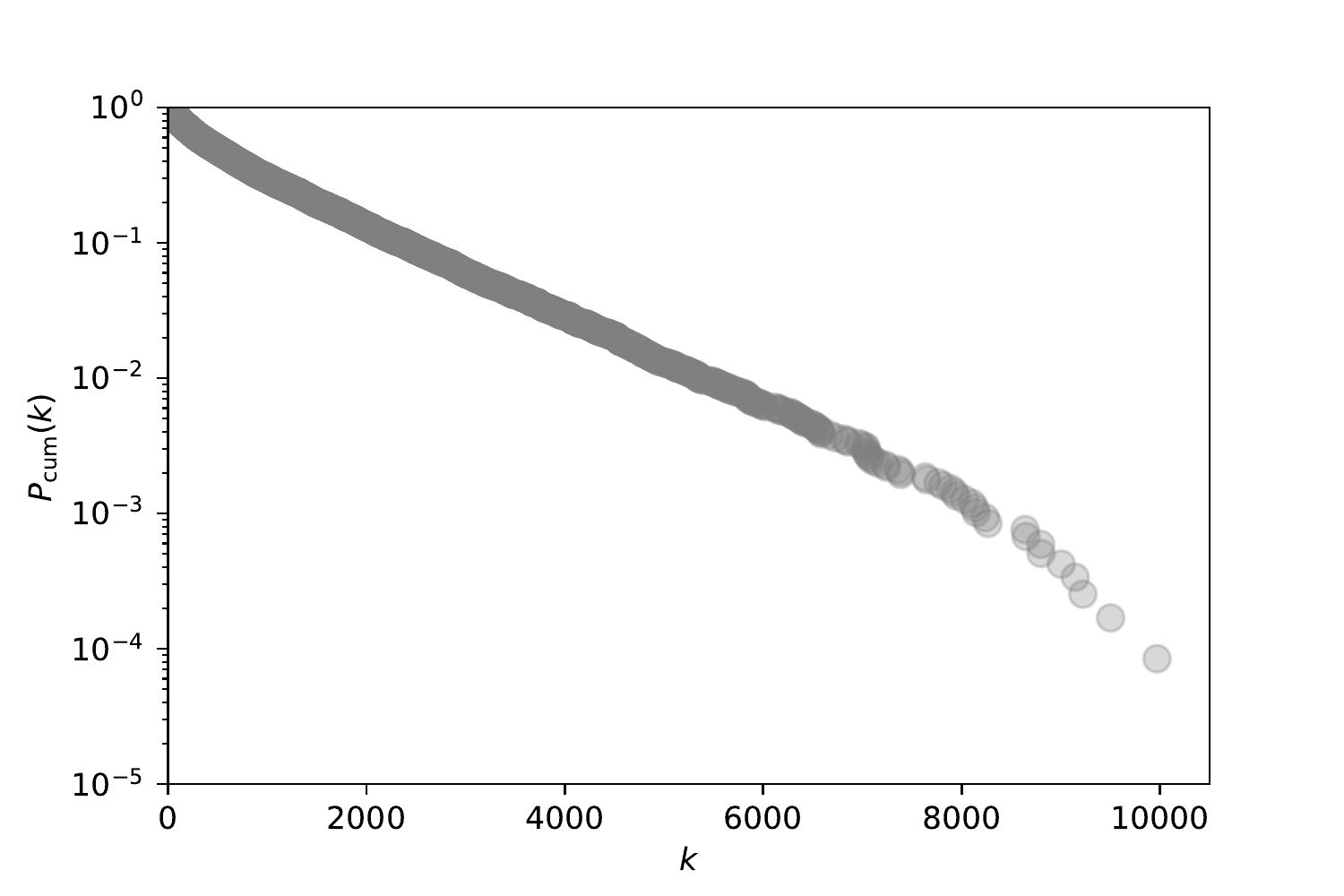}\\
 (a) & (b)
\end{tabular}
}
 \caption{Panel {\bf a}: Node degree distribution $P(k)$ in a double logarithmic scale and a power law function $k^{-\gamma}$ with the exponent $\gamma=1$ shown by a solid line for better visual orientation. Panel {\bf b}: Cumulative function  $P_{\rm cum}(k)$ in a semi-logarithmic scale.}
 \label{fig:emp_degree_distributions}
\end{figure}

The connectivity patterns of different nodes/concepts vary: rather expectedly, more specific concepts appear with a few others only, while more generic ones co-occur with a lot of others. For example, \texttt{Statistics} concept has co-occurred with $9,970$  concepts ($84\%$ of other concepts in the data set) and has the highest degree in the network $k_{\rm max}=9970$. Standard deviation of the node degree distribution  reaches value of $\sigma=1,146$ and indicates high level of inhomogeneity among concept co-occurrence statistics. The above features of node degree distribution may hint for its skewed shape. Indeed, the tail of node degree distribution $P(k)$ shown in Fig.~\ref{fig:emp_degree_distributions}{\bf a} in a double logarithmic scale may be visually compared to a power-law function $k^{-\gamma}$ with exponent close to $\gamma=1$, while cumulative degree distribution $P_{\rm cum}(k)$ may be well fitted by a straight line in semi-log scale, see Fig.~\ref{fig:emp_degree_distributions}{\bf b}, which hints for its exponential decay.
Due to the finite size of the data set under investigation the functional form of the degree distribution cannot be definitely identified. 

Another feature of the node degree distribution is the increase of $P(k)$ with $k$ for small degrees, see Fig.~\ref{fig:emp_degree_distributions}{\bf a}. It means that, e.g., the probability of a randomly selected node $i$ to have degree $k_i=3$ is smaller than its probability to have degree $k_i=30$. This is quite natural since the number of distinct concepts in a manuscript sets a lower bound for node degrees. Indeed, all concepts met in the same manuscript are represented by a complete sub-graph (or clique). Therefore, the maximum at around 30 concepts per article in Fig.~\ref{fig:1.concepts_per_article} may correspond to the maximum for the node degree distribution (see Fig.~\ref{fig:emp_degree_distributions}{\bf a}).
Similar shapes of node degree distributions were found and declared to be robust for a few other empirical data sets analysed in \cite{Cattuto2009}, where the co-usage of topical tags annotating various web resources is considered and for the network of chemical entities co-occurring in the abstracts of research papers and patents \cite{Rzhetsky2015}. The summary of basic network metrics is shown in the first line
(\texttt{empirical}) of the Table~\ref{tab:table_1}. 
\begin{table*}[!ht]
    \centering
    {\small
    \begin{tabular}{p{2.3cm}|p{0.7cm}|p{0.8cm}|p{0.7cm}|p{0.6cm}|p{0.6cm}|p{0.8cm}|p{0.7cm}|p{0.5cm}|p{0.5cm}}
         & $N$ & $L$, $\times10^6$& $\rho$, \% & $\langle{k}\rangle$ & $\sigma$ & $k_{\rm max}$ & $r$ & $\langle{c}\rangle$ & $T$ \\ \hline
        \texttt{empirical}  & 11853 & 5.38 & 7.66 & 908 & 1146 &  9970 & -0.28  &  0.74  & 0.38 \\\hline
        Erd\H{o}s-R\'enyi  & 11853 & 5.38 & 7.66 & 908 &   29 & 1023 &  0.00  &  0.08  & 0.08  \\
        Barab\'asi-Albert        & 11853 & 5.38 & 7.66 & 908 &  568 & 3875 &  0.01  &  0.15  & 0.15  \\
        \hline
        \texttt{USP}, 37    & 11551 & 14.62 & 21.92 & 2531 &  1516 & 5204 &  0.23  &  0.40  & 0.45 \\
        \texttt{USP}, EMP     & 11538 & 19.65 & 29.53 & 3407 &  1850 & 6433 &  0.19  &  0.49  & 0.53 \\
        \texttt{PSP}, 37    & 11557 &  0.86 &  1.28 &  148 &   618 & 8602 & -0.55  &  0.94  & 0.06 \\
        \texttt{PSP}, EMP     & 11554 &  1.50 &  2.25 &  260 &   788 & 7603 & -0.62  &  0.95  & 0.12 \\
    \end{tabular}}
    \caption{Basic features of concepts networks addressed in our study. The first line (\texttt{empirical}) corresponds to the empirically observed network, the rest characterise different models discussed
    in the paper. Network properties: the number of nodes $N$, number of links $L$, density of links $\rho$,
    average node degree $\langle{k}\rangle$, standard deviation of the node degree $\sigma$, maximal node degree $k_{\rm max}$,
    assortativity mixing by degrees $r$,
    average clustering coefficient $\langle{c}\rangle$ and transitivity $T$. 
    For all network models, the table contains average values of the corresponding characteristics averaged over 100 realizations (with $\nu=8.8\times10^{-3}$ for generated networks, see below for the model details).}
    \label{tab:table_1}
\end{table*}

Besides the nodes have distinct degrees, it is important to check whether the nodes tend to be connected to the nodes that are similar to them. Such node degrees correlations can be found in real networks: e.g., high-degree nodes prefer to connect to the other high-degree nodes more likely than one would expect by chance for social networks (including co-authorship networks) while the situation is the opposite for technological networks, where high-degree nodes prefer to connect to low-degree nodes \cite{newman2002assortative}. 
Such preferences may be quantitatively measured in terms of assortativity mixing by degrees $r$, 
which is defined as a Pearson correlation coefficient between node degrees \YuH{on both ends of 
each link in the network.} Positive values of $r$ reflect degree homophily (i.e. that high-degree nodes are likely to be connected to the other high-degree nodes), while negative values of $r$ indicate that high-degree nodes tend to be connected to the low-degree ones. The considered network of concepts has negative assortativity, $r=-0.28$: the similar absolute value (same order of magnitude) as for the co-authorship network ($r=0.36$, see \cite{newman2002assortative}) but opposite sign. This indicates that unlike the mechanisms in social networks, in concepts network high-degree nodes tend to attract low-degree ones. Presumably, to describe a specific problem, a list of specific terms is used, but some more generic ones are used to give a context to the wider audience or to connect specific problem/domain to more widely known existing knowledge.

Node degree, as described above, reflects the level of connectedness of a given concept to others in the network. One may also investigate the local connectivity patterns around the nodes: how densely interconnected closest neighbours of a node are. Since all concepts found in the same document are fully interconnected by definition, the lower connectivity of node's nearest neighbours indicates that the corresponding concept was used in different papers bringing different set of concepts together.
%
If given node $i$ has $k_i$ concepts/nodes connected to it by links, one may ask a question how many of these $k_i$ nodes are directly interconnected. Given there are $m_i$ such connections, the ratio between $m_i$ and the number of all such possible connections $k_i(k_i-1)/2$ is defined as a clustering coefficient $c_i$ of node $i$
\begin{equation}
    c_i = \frac{2m_i}{k_i(k_i-1)}, \hspace{2em} {\rm for } \hspace{1ex} k>1.
\end{equation}
Average clustering coefficient is defined as an average value of $c_i$
\begin{equation}
    \langle{c}\rangle = \frac{1}{N}\sum_{i=1}^{N}c_i
\end{equation}
where $i$ runs over all $N$ nodes in the network.
In general, this metrics does not correlate with the node degree or network density. These correlations rather depend on the network type. For example, average clustering coefficient $\langle{c}\rangle$ of the fully connected network or of the network that consists of isolated fully connected sub-graphs equals $1$, while for a tree-like network $\langle{c}\rangle=0$ is independent of its density. For some network models (e.g. the Erd\H{o}s-R\'enyi random graph) clustering coefficient $\langle{c}\rangle=\langle{k}\rangle/(N-1)$ by definition (for the graph of a large enough size). The reason is that with the underlying design of the graph, a probability that two randomly chosen neighbours of node $i$ have a link between each other is the same as the probability that two randomly chosen nodes are connected by a link.

An alternative way to investigate local correlations is to consider transitivity $T$ (sometimes referred to as global clustering coefficient, see \cite{barabasi2014network}). It is defined as a ratio between the number of closed triplets in the network and the total number of network triplets \cite{luce1949}. Similarly to the average clustering coefficient $\langle{c}\rangle$, $T=1$ for a fully connected network and $T=0$ for a tree-like graph. However, the difference between the two ($\langle{c}\rangle$ and $T$) may indicate some special topological features of the network. 

To get a better understanding of the processes behind generation of scientific knowledge and the corresponding knowledge graph, let us compare the resulting empirical concepts network
topology with the topology of networks of a similar size obtained within familiar generative models.

\subsection{Comparison with the Erd\H{o}s-R\'enyi model}\label{II.3}
Generation of scientific knowledge involves creativity which is not a deterministic process. Moreover, each author may use different terms to describe the same observation. This may lead to the basic assumption of random connections between concepts, which corresponds to the maximum entropy principle given no other restrictions besides number of nodes and links. In terms of networks this process is represented by the 
Erd\H{o}s-R\'enyi random graph \cite{ER,barabasi2014network}. Exponential decay of the cumulative degree distribution and the 
existence of a clear maximum of $P(k)$ support the idea behind a choice of such model.

This model allows to generate uncorrelated networks with the same number of nodes $N$ and links $L$ as the original one. To this end, choosing $N$ and $L$ as given in the first row
of Table \ref{tab:table_1} we have generated 100 realizations of the Erd\H{o}s-R\'enyi graph. Since we deal with dense networks, we found in practice that every single realization of the Erd\H{o}s-R\'enyi graph does not contain disconnected components and consists of a single connected component. More detailed description of the empirical concepts network features 
is shown in the second row of the  Table.~\ref{tab:table_1}.\footnote{For this and subsequent models, an averaging over 100 network realizations has been performed. For the network sizes explored, the self-averaging effect has been observed: all network characteristics reported in the Table coincide for each network realization with their average values within the reported accuracy. The only feature that slightly fluctuates, is the maximal node degree.} 

\begin{figure}[!ht]
 \centerline{\includegraphics[width=9.0cm]{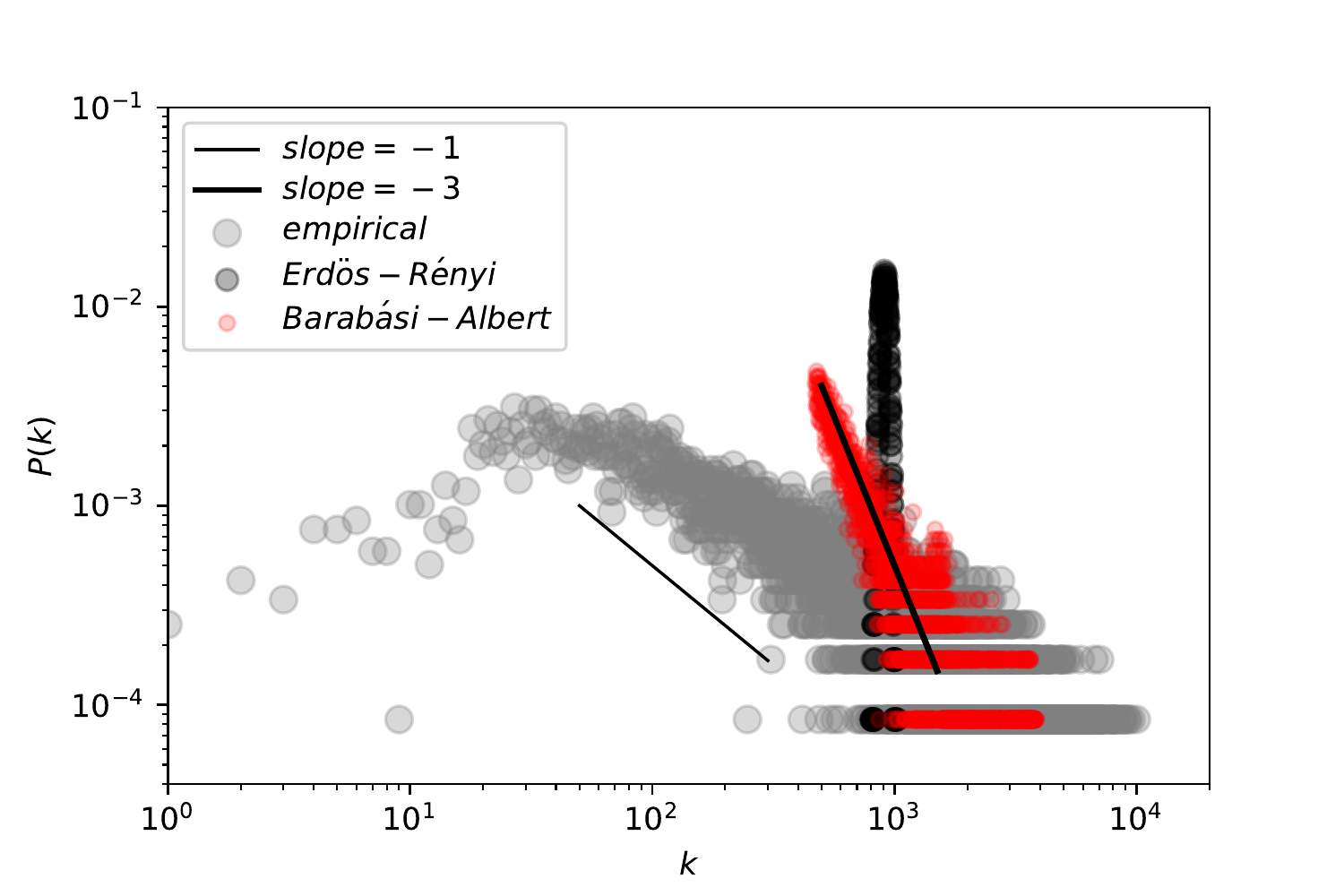}}
 \caption{Empirical node degree distribution (grey discs) is compared with the node degree
 distributions of the Erd\H{o}s-R\'enyi  (black discs) and Barab\'asi-Albert (red discs) graphs of the same  size, see the text for a whole description. Solid lines show power law functions $k^{-\gamma}$ with the exponents $\gamma=1,3$, correspondingly.}
 \label{fig:degree_distributions}
\end{figure}

Given the same number of nodes and links, the density of links $\rho$ and the average node degree $\langle{k}\rangle$ coincide in the empirically observed network and the ones generated by the Erd\H{o}s-R\'enyi model. The discrepancies become visible with more in-depth analysis. Maximally observed node degree $k_{\rm max}$ in networks generated by the Erd\H{o}s-R\'enyi model exceeds its average value $\langle{k}\rangle$ by $12\%$ only, while in the empirical concepts network $k_{\rm max}$ exceeds $\langle{k}\rangle$ by almost $1000\%$. In terms of standard deviation $\sigma$, its value for Erd\H{o}s-R\'enyi random graphs is almost $40$ times smaller than for the empirical one. These mean that the empirical concepts network is much more heterogeneous than the Erd\H{o}s-R\'enyi random graph. This may be visually observed in Fig.~\ref{fig:degree_distributions}, where beside the node degree distribution for the empirical concepts network (grey discs), the corresponding distribution for a single realization of the network generated by the Erd\H{o}s-R\'enyi model is shown by black discs.
The discrepancies between the empirical concepts network and the random graph topologies are also observed for connectivity patterns between nodes of different degrees. While the empirical  concepts network is disassortative ($r=-0.28$), random network is neither disassortative nor assortative with $r=0.00$. This indicates the nature of the model where no correlations have been incorporated during the process of network generation. Finally, the random network differs from the empirical one in terms of the clustering coefficient $\langle{c}\rangle$ and transitivity $T$. The fact that the values of both $\langle{c}\rangle$ and $T$ are smaller for the random graph than for the empirical network is not a big surprise. The most insightful observation is that $\langle{c}\rangle$ and $T$ are almost identical for the random graph, while they differ significantly for the empirical concepts network. The above observations indicate that random processes incorporated into the Erd\H{o}s-R\'enyi model are not sufficient enough to describe the processes behind creative processes of scientific writing; and more sophisticated models should be considered.

This leads us to compare the empirical concepts network with another model that could arrive at a more heterogeneous and more correlated graph than the Erd\H{o}s-R\'enyi one. Let us consider the Barab\'asi-Albert model that has
growth and preferential attachment as key ingredients and is known to generate heterogeneous graph with power law decay of the node degree distribution \cite{barabasi1999}.

\subsection{Comparison with the Barab\'asi-Albert model}\label{II.4}
To remind, this generative model starts with $m_0$ isolated nodes at time $t=0$. At each time step $t>0$ a node is added and it is connected by $m(\leq m_0)$ links to the existing nodes. The choice of which nodes to connect to is governed by the preferential attachment scenario: the more connections the existing node has, the more likely it will be selected for upcoming links. Here we consider linear preference as it has been originally proposed in \cite{barabasi1999}. This model is known to produce heterogeneous graphs that in the limit of $t\to\infty$ are characterized by a power-law node degree distribution $p(k)\sim k^{-\gamma}$ with the exponent $\gamma=3$. We expect that due to its heterogeneity the network generated by the Barab\'asi-Albert model will have more similar topology to the empirical concepts network than the Erd\H{o}s-R\'enyi random graph.

To this end, we generated networks that have the same number of nodes $N=11,853$ and only slightly different number of links as compared to the empirical concepts network.\footnote{In general, the Barab\'asi-Albert process cannot arrive at arbitrary predefined number of links without additional processes that involves link removal or addition of extra links, since at each time the links arrive at blocks of size $m$. We decided to stick to the original formulation of the model with $0.005\%$ discrepancy in the number of links instead of arriving at exactly the same number of links in cost of modification of the underlying processes.} 
This is achieved by starting with $m_0=473$ isolated nodes and adding step-by-step nodes with $m=m_0$ connections each. Doing so, after $11,380$ steps we arrived at a network with $N=11,853$ nodes and $L=5,382,740$ links. As in the former case of the Erd\H{o}s-R\'enyi model, we have generated $100$ realizations of the Barab\'asi-Albert model  and found that  due to the network size the self-averaging occurs: the characteristics of the resulting networks hardly depend on the realization, they are summarized in the third row of Table~\ref{tab:table_1}.

As expected, the Barab\'asi-Albert model reproduces empirical network topology better than the Erd\H{o}s-R\'enyi model does, especially in terms of node degree heterogeneity. For example, maximal node degree exceeds its average value by more than $300\%$ (vs $12\%$ for the Erd\H{o}s-R\'enyi model and $1000\%$ for the empirical concepts network). Regarding standard deviations, even though $\sigma$ for the Barab\'asi-Albert model is twice smaller than for the original concepts network, it exceeds its value for the Erd\H{o}s-R\'enyi graph in almost $20$ times. The difference in node degree heterogeneity is clearly visible in Fig.~\ref{fig:degree_distributions}, where node degree distributions $P(k)$ for single realizations of the Erd\H{o}s-R\'enyi graph and the Barab\'asi-Albert network are shown. It is also seen that neither of the models can reproduce the empirical node degree distribution.

Similarly to the Erd\H{o}s-R\'enyi graph, the Barab\'asi-Albert network is neither assortative, nor disassortative, indicating the feature of the empirical concepts network that cannot be captured by the model. The other feature that is not captured by the model is the difference between the average clustering coefficient $\langle{c}\rangle$ and transitivity $T$, even though the values for both are closer to the empirical concepts network than the ones for the Erd\H{o}s-R\'enyi network. 

To understand possible mechanisms that lead to the  concepts network under consideration, let us develop a model that is capable to reproduce its empirically observed features. Doing so, we will not put as a primary goal to rich a high precision of reproducing given set of metrics. Rather we will be interested in a qualitative description of main tendencies in network structure and their explanation by network
generation mechanisms. 

\section{Growth by blocks with preferential selection}\label{III}
The model we suggest to describe the concepts network growth is based on two main features: i) growth by blocks and ii) preferential selection. By growth by blocks we mean that every generated article enters the concepts network as a complete sub-graph of concepts it contains. By preferential selection we refer to the concept selection mechanism. The concepts selected to populate an article may be novel as well as already existing ones. In the case of selection from the existing concepts, the preference is given to the concepts that appeared more frequently in the past.

\subsection{Discrete time process}

Let us consider a discrete time process where time $t$ changes in a range $t=1\ldots {\cal N}$. At each time step $t$ a new article is generated. We will refer to this article by $A_t$. Each article $A_t$ is modelled as a set of $n_t$ distinct concepts, i.e. a block of $n_t$ concepts. The process of article generation consists of i) drawing the block size $n_t$ and ii) selection of $n_t$ concepts to populate the article, see Fig.~\ref{fig:model_demonstration} for the process demonstration. Below we describe the details of $n_t$ choice and the concepts selection procedures.

\subsection{Block sizes}
In order to correspond to the empirically observed network,  we will ensure the average block size $n_t$ (i.e. the average number of concepts per article) to be the same as in the empirical data set. Here, we consider two alternative ways to satisfy this condition. At the first instance, we employ the simplest mechanism such that $n_t=37$ is fixed and independent of $t$. \textcolor{black}{This strategy can be found, e.g., in the generative models proposed in \cite{Rzhetsky2015} (the network of co-used chemical annotations is growing by adding a link on each time step, i.e., $n_t=2$) and \cite{Wang2010} (hyperedge containing three nodes is repetitively added, i.e., $n_t=3$). }
Second, we draw $n_t$ from the actual distribution of the number of concepts per article in the \texttt{arXiv} data set. 
An example is given in the first row of Fig.~\ref{fig:model_demonstration}, with 
$n_1=4$ (four concepts $C_1$, $C_2$, $C_3$, $C_4$ in article $A_1$), $n_2=3$ (three concepts $C_1$, $C_5$, $C_6$ in article $A_2$) and $n_3=2$ (two concepts $C_3$, $C_4$ in article $A_3$).
Such settings allow us to analyze whether the variations in block sizes affect the resulting network topology.

Once $n_t$ is defined, the next step is to select the concepts themselves. The following subsection describes the mechanisms that are used in the model to perform such selection.

\subsection{Selecting the concepts set}
\begin{figure}[!ht]
 \centerline{\includegraphics[width=1\textwidth]{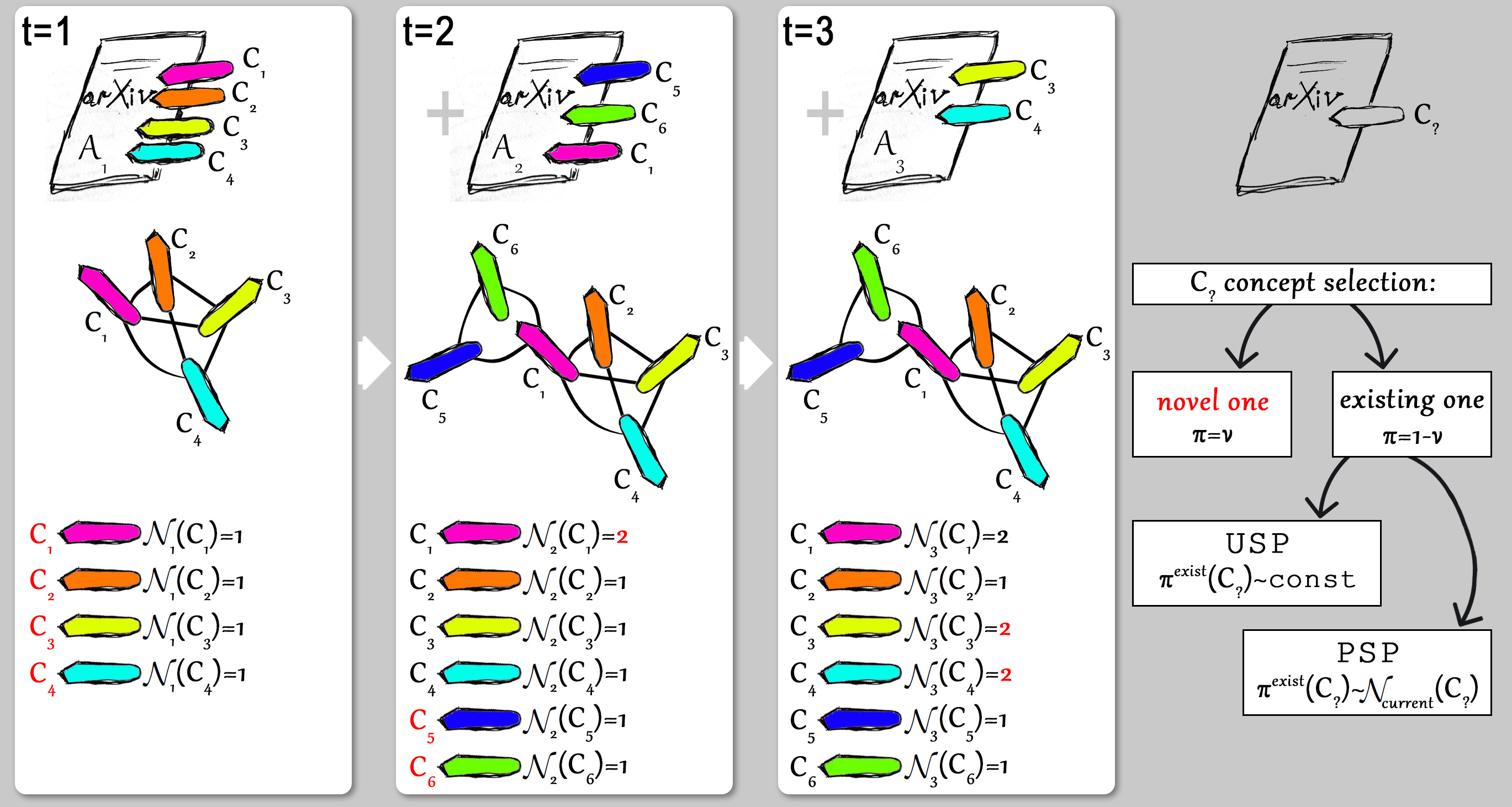}}
 \caption{Model demonstration. At each time step $t$ a new article is generated. An article $A_t$ containing $n_t$ concepts enters the concepts network as a complete graph, and affects the number of occurrences ${\cal N}_{t}(C_i)$ of a concept $C_i$. Once a new article $A_t$ is being generated, all $n_t$ concepts slots should be filled out in a probabilistic way. A concept $C_i$ may be either a \emph{novel} one (the one introduced in the article $A_t$ for the first time), or one the concepts that appeared on previous time steps. The model considers two scenarios for \emph{existing} concept selection: uniform selection process (\texttt{USP}) and preferential selection process (\texttt{PSP}).}
 \label{fig:model_demonstration}
\end{figure}

Let us consider a time step $t>1$.  The generated up to this moment data set consists of $t-1$ articles and $N_{t-1}$ different concepts (let us denote the set of these concepts by $\mathbb{C}_{t-1}$), see Fig.~\ref{fig:model_demonstration} for illustration.
The article $A_t$ generated at time $t$ may contain some of the above $N_{t-1}$ concepts as well as the concepts that are
introduced in the article $A_t$ for the first time. Let us call the latter ones as \texttt{novel} concepts. Within our
model we fix the probability  $\nu$ for each of the article concepts to be a novel one. Consequently, a concept of 
the  generated article is one of already existing $N_{t-1}$ concepts with probability 
$1-\nu$. Moreover, let us consider that different already existing concept have different chances to be selected 
to populate an article: the more popular the concepts is among first $t-1$ articles, the more likely it will be 
selected to populate the $t$-th one. We will call such process a preferential selection. For comparison, we will 
also consider a uniform selection process that picks existing concepts for article $A_t$ independently of their 
frequencies of appearances among $t-1$ first articles. Below, we describe the concepts selection mechanism in more detail.

\subsubsection{Novel vs already existing concepts}
The process of concept selection for $A_t$ is done one-by-one until the required number $n_t$ is reached. Let us describe below a procedure of selecting the $i$-th concept for article $A_t$, thus $i$ runs in a range $i=1\ldots n_t$. 

As explained above, the model says that with a fixed (quite low) 
probability 
\begin{equation}
    \pi^{\rm novel}_{t,i} = \nu
\end{equation}
the $i$-th concept is a novel one, it does not belong to the $\mathbb{C}_{t-1}$ set. Accordingly, with the probability
\begin{equation}
    \pi^{\rm exist}_{t,i} = 1-\nu
\end{equation}
the $i$-th concept belongs the concepts set $\mathbb{C}_{t-1}$. For convenience, let us denote the subset of existing concepts $\mathbb{C}_{t-1}$ contained within $t-1$ first articles and excluding $i-1$ selected concepts for article $A_t$ by $\mathbb{C}_{t\backslash i-1}$. Once the already existing concept has been selected, it cannot be selected again for the same article. Therefore it is to be selected from the set $\mathbb{C}_{t\backslash i-1}$.

Below we suggest two scenarios to select the $i$-th concept for article $A_t$ from the set of already existing concepts.

\subsubsection{Uniform selection process}
First, let us consider a uniform selection process (\texttt{USP}). This is relatively simple scenario. It assumes that each concept that appeared within $\mathbb{C}_{t-1}$ set has the same chance to be selected. I.e. the probability $\pi^{\rm exist}_{t,i}(C_j)$ that the concept $C_j$ will be selected as a $i$-th one for the article $A_t$ follows
\begin{equation}
    \pi^{\rm exist, USP}_{t,i}(C_j) =  \frac{(1-\nu)}{|\mathbb{C}_{t\backslash i-1}|}, \hspace{1em} C_j \in \mathbb{C}_{t\backslash i-1}.
\end{equation}
Here $|\mathbb{C}_{t\backslash i-1}|$ is the {\em size} of a set  $\mathbb{C}_{t\backslash i-1}$.

\subsubsection{Preferential selection process}
Second, we consider preferential selection process (\texttt{PSP}). 
Unlike in \cite{Rzhetsky2015,Iacopini2018,Cattuto2009}, the preference is not governed by the properties of the underlying network. No input information about the connectivity between concepts is required.
In this scenario, the probability $\pi^{\rm exist}_{t,i}(C_j)$ for the concept $C_j$ to be selected is proportional to the number of articles ${\cal N}_{t-1}(C_j)$ in which the concept has appeared:
\begin{equation}
    \pi^{\rm exist, PSP}_{t,i}(C_j) = \frac{(1-\nu)({\cal N}_{t-1}(C_j))}{\sum_{\ell} {\cal N}_{t-1}(C_\ell)}, \hspace{1em} C_j\in \mathbb{C}_{t\backslash i-1}\,,
\end{equation}
where the denominator sums the number of times each concept $C_\ell$ from the set $\mathbb{C}_{t\backslash i-1}$ has appeared in all articles.

In both scenarios the concepts network grows by adding cliques to the existing graph, see the third row of Fig.~\ref{fig:model_demonstration}. At each time $t$ once a new article $A_t$ of $n_t$ concepts is generated, it enters the concepts network as a complete graph of $n_t$ nodes and $n_t(n_t-1)/2$ links between them. Thus, during its evolution, the following processes may be observed in a generated concepts network: i) addition of new nodes, ii) appearance of links between the novel nodes as well as between novel and already existing nodes, iii) appearance of new links between previously unconnected existing nodes, which is important for generation of dense networks.

Although majority of complex networks structures of either natural or man-made origin are sparse, there exists ongoing interest in the so-called dense networks, that are characterized by diverging mean node degree. For the scale-free distributions this means that the corresponding decay exponent takes values between 1 and 2 \cite{Borgs08,Courtney18}.  
Examples of such structures are found in  the brain \cite{Bonifazi19},  internet 
(see \cite{Allaei06} and references therein), social recommender systems \cite{Zhou11}.
Familiar preferential attachment-based growth models are not able to reproduce properties of dense networks. 
The reason is due to the fact that the growth in such models occurs by homogeneous addition of network nodes and links that leads to sparse structures. 
Certain scenarios circumvent such restriction and lead to dense networks 
 \cite{Diaconis07,Borgs08,Wolfe13,Crane16,Caron17,Courtney18}. 
Although the observed concepts network is characterised by a very high density of links (cf. $\rho$ in Table \ref{tab:table_1}),
strictly speaking it cannot be named the dense network in the sense explained above. However, we believe that the
model suggested above for its evolution may be useful in studying other networks with high density of links.
Once the processes that govern generation of concepts network have been defined, let us explore topological properties of the network generated by the rules described above.

\section{Results}\label{IV}
In our simulations we set the number of generated articles to be exactly the same as the number of articles ($36,386$) in the \texttt{arXiv} data set. Fixing the number of articles ${\cal N}$ does not guarantee that the generated network will have the same number of nodes $N$ (concepts) as the empirical concepts network: the two remaining degrees of freedom of the model may affect the number of different concepts in a generated data set.

\subsection{Number of nodes}

\begin{figure}[ht]
 \centerline{
\begin{tabular}{cc}
\includegraphics[width=6cm]{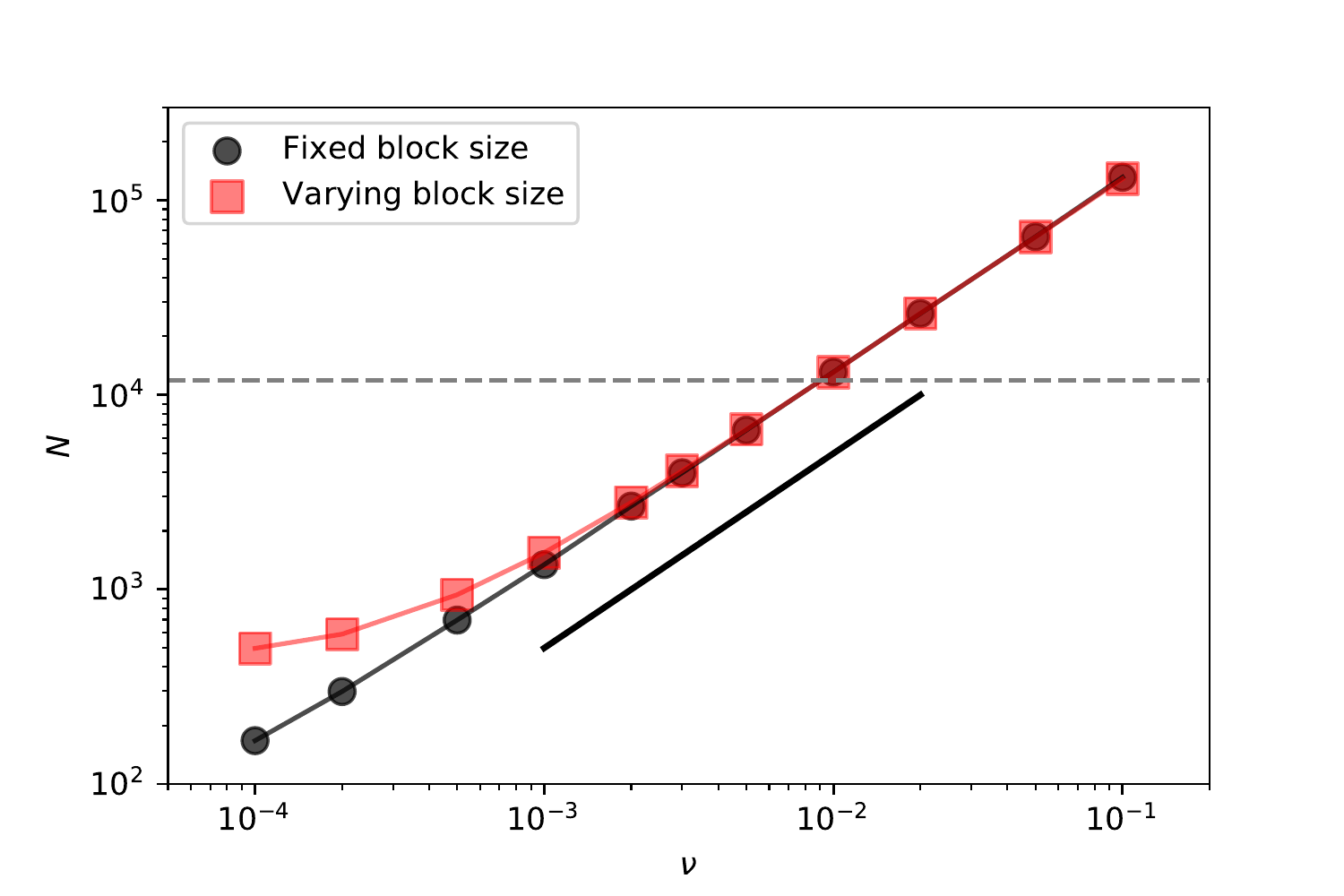} &
 \includegraphics[width=6cm]{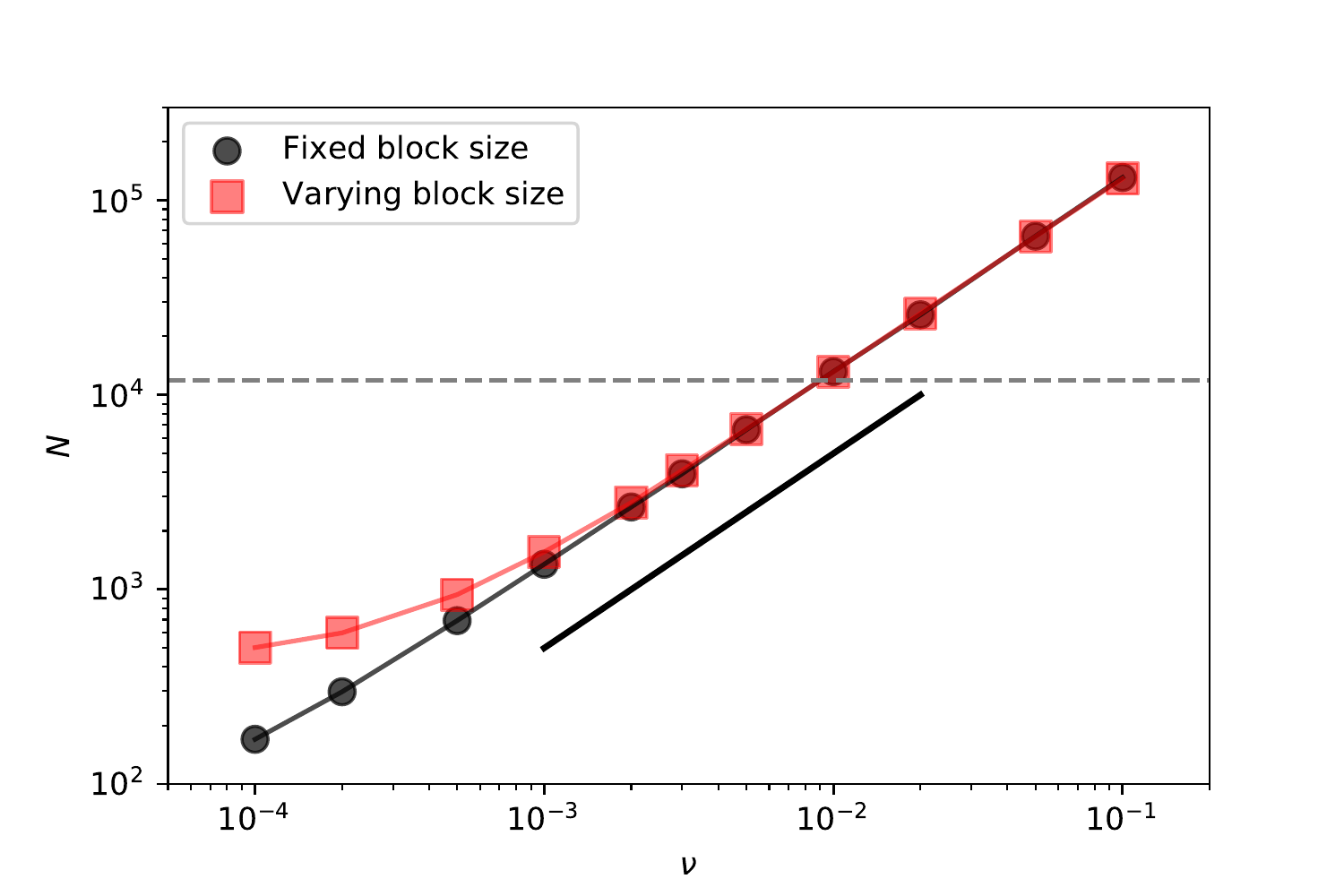}\\
 (a) & (b)
\end{tabular}
}
\caption{Dependency of the number of concepts ($N$) in the generated data sets on the probability of novel concept appearance $\nu$ for different distributions of the block sizes: fixed block size (black circles) and empirically taken varying block sizes (red squares). Panel {\bf a} corresponds to the preferential selection mechanism, while panel {\bf b} corresponds to uniform selection mechanism. All points represent average values over a number of realizations that vary from a single realization for large values of $\nu$ to 100 realizations for small $\nu$ values. Solid lines in both panels represent linear dependency and are shown for visual orientation. Dashed lines show the number of concepts $N=11853$ in the empirical data set.}
 \label{fig:2.32}
\end{figure}

It is natural to expect positive correlations between the parameter $\nu$ (the probability of appearance of a novel concept) and  the number of different concepts, i.e. the number of nodes $N$ in the resulting network. Indeed, we may see this dependency for generated data sets in Fig.~\ref{fig:2.32}{\bf a} (\texttt{PSP}) and in Fig.~\ref{fig:2.32}{\bf b} (\texttt{USP}) for different values of the model degrees of freedom: the probability $\nu$ the block size distribution.
The number of concepts in a generated data set has a tendency to increase independent of the acquired concepts selection mechanism and independent of the block size distribution. Moreover, the number of nodes looks to be rather independent on the concepts selection mechanism. For relatively large values of $\nu$ ($\nu>3\times10^{-3}$) the distribution of block sizes considered does not affect the number of concepts in the resulting data set. Thus, further in the analysis of the generated concepts networks we will fix the parameter $\nu=8.8\times10^{-3}$ that gives us approximately the same number of concepts as in the original data set for both types of block size distributions and for the two considered concepts selection mechanisms. Indeed, we performed 100 simulation runs for each set of model parameters (degrees of freedom: $\nu=8.8\times10^{-3}$; type of block size distribution; \texttt{PSP} vs \texttt{USP} mechanisms of concepts selection) and observed that the affect of the block size distribution has the same order of magnitude as the stochastic effects of the modelled process confirming our visual conclusions.
The number of concept nodes $N$ the resulting generated networks contain varies between $N=11,257$ and $N=11,863$ over all realizations. The reason to arrive at about the same numbers of concepts for \texttt{USP} and \texttt{PSP} mechanisms may be explained as follows. The difference between the two models consists in which concepts out of existing ones will be selected to populate an article. But the mechanisms to select existing or novel concepts are identical in both processes, leading to the same number of concepts in the large network limit. 
However, we continue to use both distributions of block sizes for modelling -- the difference becomes visible further.

It is worth mentioning that for small $\nu$ the number of concepts $N$ in the generated data sets depends significantly on the type of block size distribution. If $\nu$ is small enough, the number of concepts in the generated data set using empirical block sizes exceeds the number of concepts in the generated data set using fixed block sizes. This discrepancy may be explained by considering the limiting scenario with $\nu=0$. Within this scenario the number of different concepts in the generated data set equals to the maximal number of concepts observed in an article (block size $n_t$) $N = {\rm max}(n_t)$ over all articles $A_t$ in the generated data set. If each article has $n_t = \langle{n_t}\rangle$ concepts, i.e. fixed block size case, then $N = \langle{n_t}\rangle$. However, in the case of the empirical block sizes, $N$ will be higher than $\langle{n_t}\rangle$ if the distribution of $n_t$ values has a non-zero probability for at least two values.

\subsection {Number of links}
So far we adjusted model parameters to make the number of nodes in the generated concepts networks comparable to the number of nodes in the empirical concepts network. There are no parameters left to adjust the number of links for the generated network to contain. Let us now focus on the number of links found in the generated networks.

The numbers of links in the generated concepts networks have the same order of magnitude as the empirical one (millions of links). More detailed inspection, see Table~\ref{tab:table_1} shows that the number of links in the networks generated by \texttt{USP} model exceeds the number of links in the empirical concepts network in about $3-4$ times while this number in the networks generated by \texttt{PSP} is about $3-6$ times smaller than the number in the empirical concepts network.
The reason to have significantly smaller number of links in \texttt{PSP} generated network as compared to \texttt{USP} generated network is the following. In frames of the \texttt{PSP} process, a pair of ``popular'' concepts have relatively high chance to appear in every subsequent article, and such co-appearance does not introduce a new link to the network after the first co-appearance. In \texttt{USP} mechanism popularity of the concepts does not affect their chances to co-occur and new links are introduced with higher probability. Now, let us now proceed with the topological features of the generated networks.

\subsection{Node degrees}

Let us analyze the node degree distributions first. Since the number of links $L$ has a direct relation to the average node degree $\langle{k}\rangle$, average node degrees in the networks generated using \texttt{PSP} mechanism is smaller than in the empirical concepts network, while average node degrees in the networks generated using \texttt{USP} mechanism overestimate the corresponding value for the empirical concepts network, see Table~\ref{tab:table_1}. Among the two distributions of block sizes, the empirical one arrives at a closer value of the average degree if \texttt{PSP} mechanism is used. But in the case of \texttt{USP} mechanism, generated networks with fixed block sizes have a closer value of  $\langle{k}\rangle$ to the original concepts network.

\begin{figure}[ht]
 \centerline{
\begin{tabular}{cc}
\includegraphics[width=6.5cm]{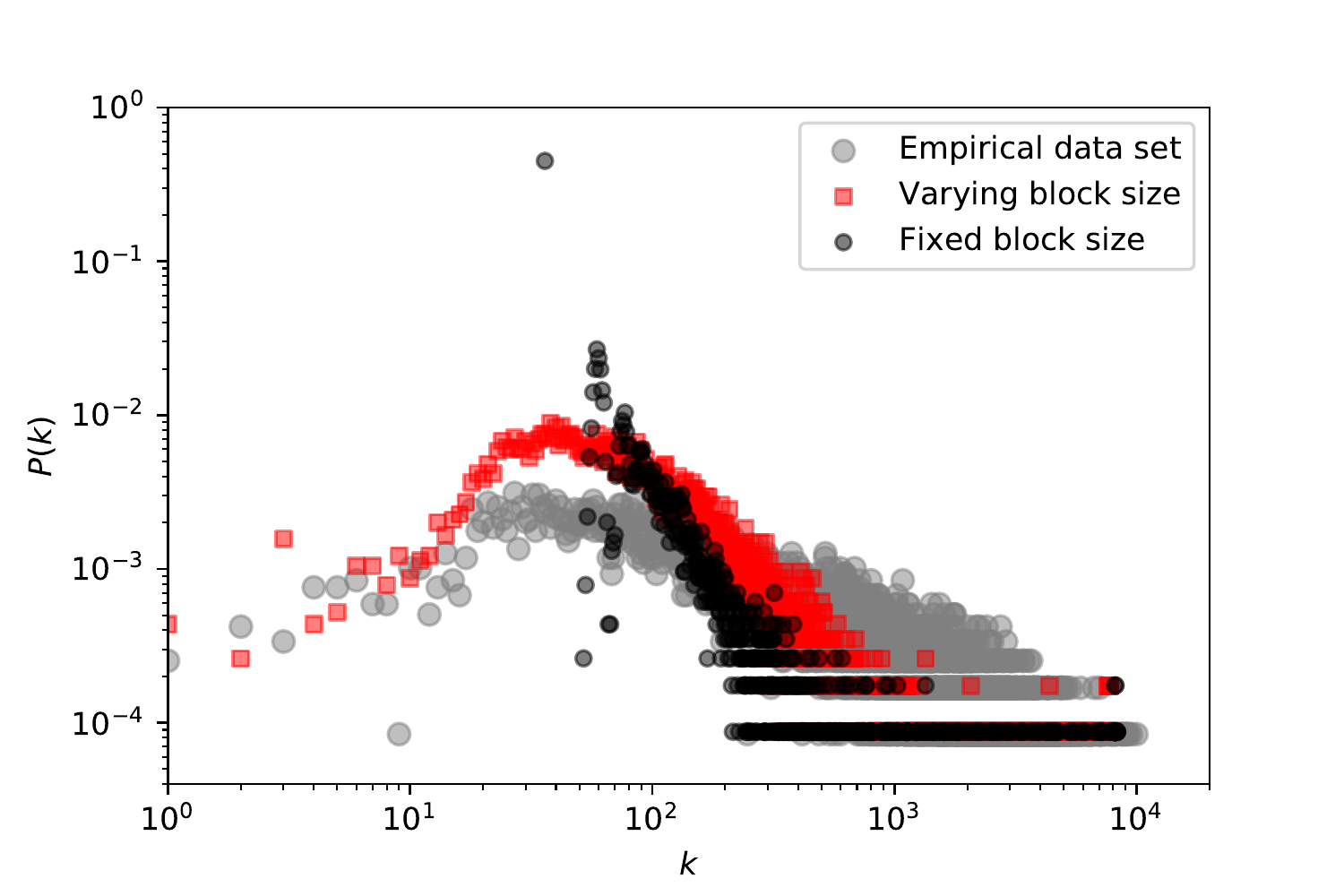} &
 \includegraphics[width=6.5cm]{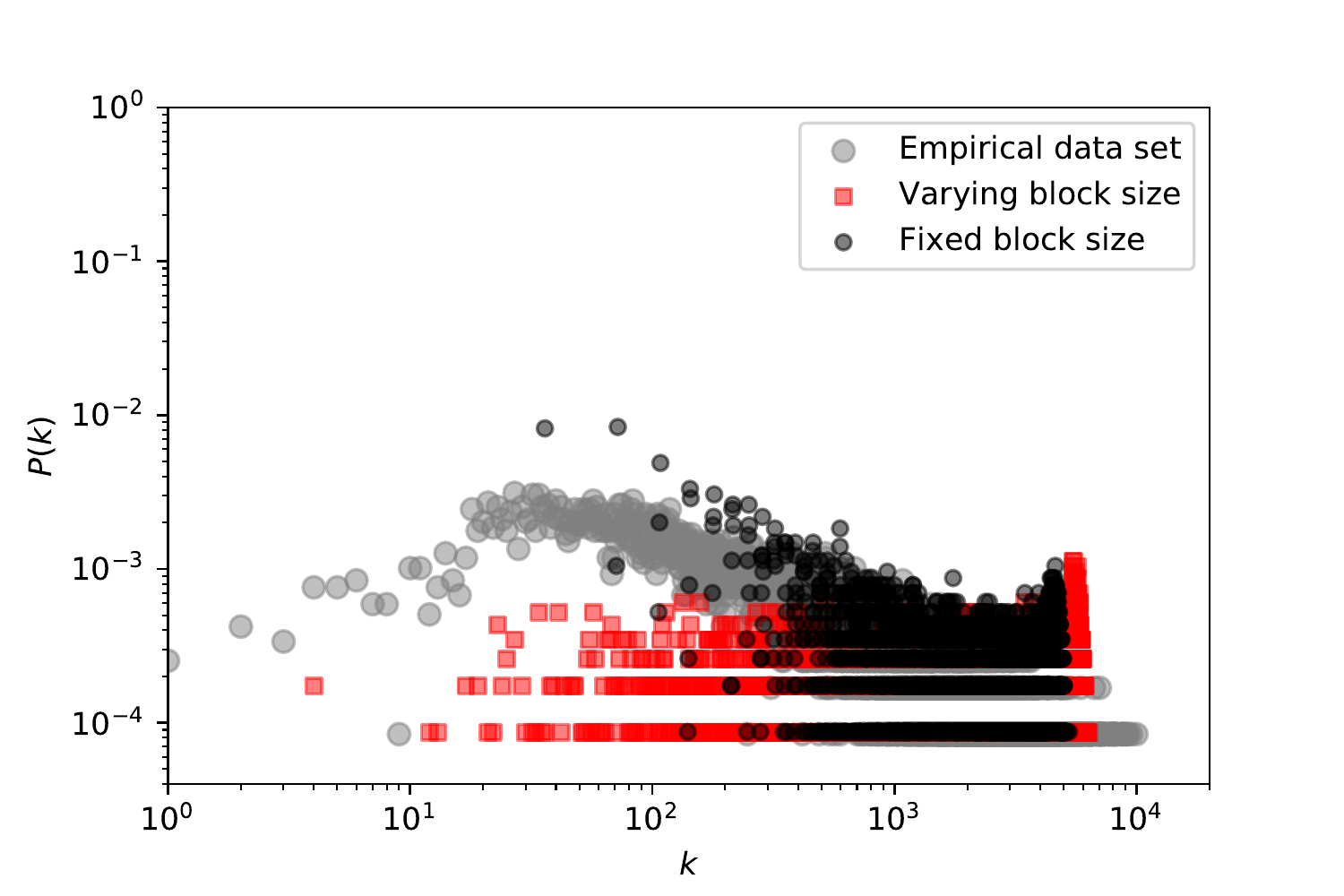}\\
 (a) & (b)
\end{tabular}
}
 \caption{Node degree distributions for \texttt{PSP} (panel {\bf a}) and \texttt{USP} (panel {\bf b}) generated networks in a double logarithmic scale. Black circles represent concepts networks generated using fixed block sizes, and red squares represent concepts network generated using varying block sizes taken from the empirical data. For comparison the picture also contains degree distributions of the empirical concepts network displayed by the grey filled circles. The results correspond to a single realization of the generated concepts networks for each set of degrees of freedom.}
 \label{fig:generated_degree_distributions}
\end{figure}

Node degree distributions for the generated networks using \texttt{PSP} and  
\texttt{USP} mechanisms are shown in Fig.~\ref{fig:generated_degree_distributions} (panels {\bf a} and {\bf b}, correspondingly). 
There, the distribution $P(k)$ for  the empirical concepts network is provided for comparison. 
It can be seen in the figure, that any considered concepts selection mechanisms in combination with considered block size distributions cannot precisely reproduce degree distribution of the empirical concepts network. However, it is possible to capture its general shape if \texttt{PSP} mechanism is chosen.

If \texttt{PSP} mechanism is considered in combination with the fixed block size (black discs in Fig.~\ref{fig:generated_degree_distributions}{\bf a}), degree distribution for relatively large degrees has rather a power-law decay $P(k)\sim k^{-\gamma}$ with the exponent close to $\gamma\approx3$. The power law decay with $\gamma=3$ is also  expected for the Barab\'asi-Albert network \cite{barabasi1999}, which used preferential attachment during the network generation process. The generated concepts networks has zero probability of finding a node with degree $k<36$ by construction, since the minimal block size sets a threshold for the node degree.

If \texttt{PSP} mechanism is considered in combination with varying block size (red discs in Fig.~\ref{fig:generated_degree_distributions}{\bf a}), the node degree distribution of a generated network is closer to the node degree distribution of the empirical concepts network than that for a generated concepts network with the fixed block size. Since the minimal block may consist of a single concept, there is a non-zero probability of finding an isolated node in the generated network, which is not seen in Fig.~\ref{fig:generated_degree_distributions}{\bf a} due to the logarithmic scale. Then $P(k)$ has a tendency to increase with $k$ up to $k=30\div50$ similarly to the degree distribution of the empirical concepts network. The further decay of the node degree distribution is slower than in the generated concepts network with the fixed block size, however it is faster than in the empirical concepts network. Even though \texttt{PSP} mechanism in combination with varying block size distribution does not allow to generate a concepts network that accurately reproduces node degree distribution of the empirical concepts network, it allows to generate a network with $P(k)$ that shares similar properties with the empirical network: an increase of $P(k)$ for small values $k$ and relative slow decay for large $k$ values.

\subsection{Check of other network parameters}

While node degree distribution is one of the basic parameters which describe the network, a number of other network characteristics have to be checked to get more information about its structure and the nature of data it reflects. 

\emph{Assortativity mixing by degree}. The tendency of nodes with high degree to be connected with poorly connected nodes for real network is reproduced only if \texttt{PSP} scenario of concepts selection is used for simulation, see Table~\ref{tab:table_1}. Although the absolute value of $r$ in this case is larger than the corresponding value for the empirical
network, its sign is considered as a significant indicator supporting the generative mechanism implemented. Following the interpretations discussed before, the concepts of different levels of generality are naturally combined to indicate the general topical area and more specific context. Since any topical classification is usually characterized by hierarchical structure (i.e., more specific topics belong to more general topical areas), co-usage of more general concepts and more specific ones is considered as natural.

\emph{Transitivity and mean clustering coefficient.}
Let us remind the feature of the empirical concepts network
which was not captured by the Erd\H{o}s-R\'enyi and by the Barab\'asi-Albert models. 
It is a difference between the mean clustering coefficient $\langle{c}\rangle$ and the transitivity $T$.
Our simulations show that the generative model with a uniform concept selection mechanism is unable to reproduce this difference too. However, if the preferential selection of concepts is used, the average clustering coefficient $\langle{c}\rangle$ of the generated network differs from its transitivity and is significantly higher than $T$ (see Table 1). Even though the model overestimates this difference, the considered scenario enables one to reproduce this feature qualitatively. This result highlights the importance of the preferential selection process in the proposed model.


\section{Conclusions and Outlook}\label{V}

In the first part of this work, analysis of the network of scientific concepts built on real data is performed.  A number of specific features such as high density, disassortativity, difference between transitivity and mean clustering coefficient together with a skewed node degree distribution were found. In the second part of this work, attempts to find an appropriate model to reproduce such combination of network features were made. It was shown that commonly used network models --  the Erd\H{o}s-R\'enyi graph and the Barab\'asi-Albert model -- fail to generate a network with the desired properties. Therefore, we have proposed a simple generative model. It is based on the general logic of scientific concepts usage: the concepts do not arrive in isolation, but a group of them has to be used to describe the content. Therefore, it is natural to model network growth as an arrival of sets of concepts with each new paper. In this case,  a fully-connected group of nodes represents this set of concepts. 

A particular feature of the empirical network of scientific concepts is the high 
value of links  density. The value of  $\rho$  found in our study means that 
any randomly chosen concept co-occurs on average with any other concept with 
the probability $\rho\simeq7.66\%$. In turn, this shows that scientific concepts 
are densely connected within the considered discipline. The proposed generative model 
reproduces and explains this network feature. Although the resulting network strictly 
speaking does not belong to the class of dense networks actively discussed in the 
literature  \cite{Diaconis07,Borgs08,Wolfe13,Crane16,Caron17,Courtney18}, the 
model suggested for its evolution may be useful in studying other networks 
with high density of links.

Growth of the concepts network happens not only by adding new nodes and attaching them to the existing ones in the graph, but also by the emergence of new links between the previously existing nodes. The latter case may correspond to the appearance of links between the 
established scientific fields and may refer to atypical combinations of scientific knowledge \cite{Uzzi2013}. As we have shown, the two mechanisms have to be taken into account to
get satisfactory results in modeling such phenomena: i) {\em growth by blocks} and ii) {\it preferential selection} of concepts. The proposed simple model allows one to generate a network with properties qualitatively similar to the properties of the empirical concepts network. Neither of these mechanisms on its own gives a satisfactory outcome: the observed structure of the network is reproduced due to their interplay.

The  discussions around topics of scientific discovery, appearing of novelty, disruptive and developing character of research and many others are not new but they are still ongoing \cite{Kuhn1977,Uzzi2013,Rzhetsky2015,Thurner2020}. However, there are no much possibilities to provide quantitative description of these processes. To our believe, the formalized network of scientific concepts allows one to highlight one of many aspects of knowledge development. Therefore, our case study of empirical data as well as the model suggested here enrich the toolkit for quantitative study of the process of scientific research.

Another possible continuation of our analysis may be achieved by
application of the hypernetwork framework, when
the notion of a relation between two objects is generalized 
 to relations between many objects \cite{johnson2013hypernetworks}.
Such approach is effective in modeling processes when ``bags of entities'' 
arrive to join the existing network and is used to describe, e.g.,  the network of authors, citations or references joined by paper or journal, see \cite{Wang2010,Hu2021}.

\section*{\YuH{Acknowledgements}}

This work was supported in part by the National Academy of Sciences of Ukraine, project KPKBK 6541030 (O.M.
\& Yu.H) and by the National Research Foundation of Ukraine, project 2020.01/0338 (M.K.).

\YuH{
\section*{Appendix}\label{A}

In this appendix we give  some examples of generic and non-generic concepts
extracted by the \texttt{ScienceWISE.info} platform from several papers
submitted to the e-print repository \texttt{arXiv.org}.\footnote{\texttt{ScienceWISE.info} web-page 
accessed July 26, 2021.}

\begin{itemize} 
    \item Albert-L{\'a}szl{\'o}  Barab{\'a}si,  R{\'e}ka   Albert.   Emergence of scaling in random networks. arXiv:cond-mat/9910332 [cond-mat.dis-nn] \cite{barabasi1999}.

{\em Generic concepts}:  Networks, Topology, Probability, Communication, Field, Generic property.

{\em Non-generic concepts}:  Scale-free, Preferential attachment, Complex systems, Graph, Random graph, Neural network, Numerical simulation, Poisson distribution, Protein, Scale invariance.

\item   V. Palchykov, M. Krasnytska, O. Mryglod, Yu. Holovatch. A mechanism for evolution of the physical concepts network. 	arXiv:2106.01022 [physics.soc-ph] \cite{Palchykov21}.

{\em Generic concepts}:  Networks, Picture, Probability, Precision, Simulations, Topology.

{\em Non-generic concepts}: Complex network, Graph, Barabasi-Albert model, Clustering coefficient, Ontology, Bipartite network, Degree distribution, Preferential attachment, Random graph, Standard deviation, Primary, Assortative mixing, Complex systems, Key phrase, Pearson's correlation, Semantic network, Statistics.

\item  M. Krasnytska, B. Berche, Yu. Holovatch, R. Kenna. Ising model with variable spin/agent strengths. arXiv:2004.05134 [cond-mat.stat-mech] \cite{Krasnytska20}.

{\em Generic concepts}:  Spin, Networks, Free energy, Geometry, Temperature, Thermodynamic limit, Topology, Magnetic moment, Particles, Probability, Thermodynamics,  Magnetic field, Symmetry.

{\em Non-generic concepts}: Ising model, Phase diagram, Statistical physics, Universality class, Many-body systems, Partition function, Graph, Scale-free,   Quenching, Critical exponent, Degree distribution, Disorder, Exact solution,   Phase transitions, Random graph, Scaling law,  Adjacency matrix,   Concurrence, Continuous Spin, Degree of freedom, Duality, Hamiltonian, Magnetization,  Mean field, Mean-field approximation, Network model, Polydispersity, Potts model, Social network, Social systems, Spin glass, Spontaneous magnetization.
    \end{itemize}
}

\end{document}